# A Survey of Incentives and Mechanism Design for Human Computation Systems

## Yuan Liu[1,2] and Chunyan Miao[1,+]

[1]Joint NTU-UBC Research Centre of Excellence in Active Living for the Elderly (LILY), Nanyang Technological University, Singapore 639798 [2]Software College, Northeastern University, China

## ABSTRACT

Human computation systems (HCSs) have been widely adopted in various domains. Their goal is to harness human intelligence to solve computational problems that are beyond the capability of modern computers. One of the most challenging problems in HCSs is how to incentivize a broad range of users to participate in the system and make high efforts. This article surveys the field of HCSs from the perspective of incentives and mechanism design. We first review state-of-the-art HCSs, focusing on how incentives are provided to users. We then use mechanism design to theoretically analyze different incentives. We survey the mechanisms derived from state-of-the-art HCSs as well as classic mechanisms that have been used in HCSs. Finally, we discuss eight promising research directions for designing incentives in HCSs.

## 1 Introduction

Over the last decades, we have witnessed significant advances in computers' ability to solve complex problems. Once restricted to tasks that were too tedious for humans, computers can now tackle a range of problems that require human-like intelligence. However, there are still many problems that the human brain can easily solve but that computers cannot. In these instances, an interesting solution is to combine humans and computers in a hybrid system – the computer system relies on the intelligence of humans to solve computational problems that it cannot solve efficiently (or with acceptable accuracy).[1–3] This paradigm is called *human computation*, and the resulting system is called a *human computation system* (HCS).[4–10] In the last decade, the widespread adoption of online communication technologies has provided an essential component for this kind of systems: the ability to easily and quickly harvest the brain power of a large number of human users. As such, numerous HCSs have been successfully developed and implemented during that time frame. Well-known examples include Yahoo! Answers, Wikipedia, and Amazon Mechanical Turk. In a short time span, HCSs have attracted hundreds of millions of users. For example, Yahoo! Answers alone hit 200 million users worldwide in 2009.[11] We can therefore conclude that such systems fill a gap and are important for a large number of people. It should not be surprising then that the research community has shown a growing interest for this field. A search for "human computation" in Google Scholar yields more than 4000 results for the past 10 years, and the number of papers has been steadily increasing year after year during that time frame.

Due to the recent emergence of this field as well as the fast growth it has experienced, there is currently no commonly agreed upon definition of *human computation* in the computer science literature. Different researchers therefore adopt different definitions, depending on their particular purposes. In,[1] human computation is defined as *"a paradigm for utilizing human processing power to solve problems that computers cannot yet solve."* In,[12] their definition of human computation comprises two important elements: 1) *"the problems fit the general paradigm of computation, and as such might someday be solvable by computers"*, and 2) *"the human participation is directed by the computational system or process."* In this survey, we define a human computation system as a system of human users and computers working together to solve problems that no known computer algorithms can solve efficiently but that can be solved easily by humans.

Of course, human users play a central role in any HCS. Systems must thus be carefully designed to account for their behaviour. Human users are generally self-interested, which implies that proper incentives should be provided so as to encourage them to meaningfully contribute to a given HCS. The provided incentives will indeed affect users' decision on whether to participate in a HCS and the amount of effort they are willing to make. High participation from users that contribute their best efforts is very often the key for a HCS to produce a useful output.[3, 13–15] Given its importance in the sucess of a HCS, the incentive issue has attracted significant interest in the last decade. A search for ("human computation" AND incentive*) in Google Scholar from 2005 onwards yields 1324 results. Fig. 1 shows the ratio of human computation papers mentioning incentives over the total number of papers on human computation, by year. As can be observed, over the last 5 years, over a third of the human computation literature has also been interested in incentives, demonstrating the importance of this topic

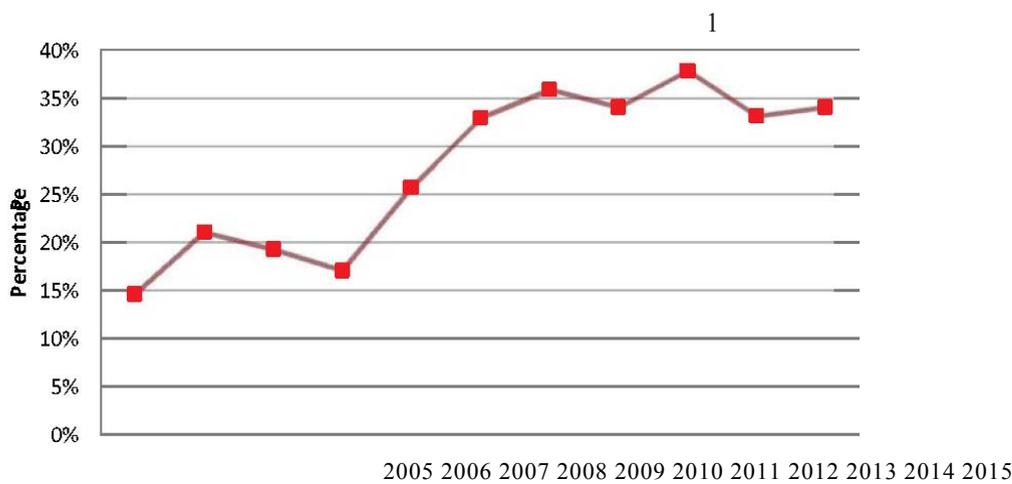

**Figure 1.** The percentage of Google Scholar results containing the keyword incentive* within all results containing the keyword "human computation", by year

in HCS research. Due to its importance and the interest it has attracted recently, this survey specifically focuses on incentive design in HCSs.

There exist two main methodologies to design incentives in HCSs: user-centered design (UCD)[16] and incentive-centered design (ICD).[17] UCD is a process in which end users' needs, wants and limitations are taken into account at each stage when designing a product or service. It is based on standards (e.g., ISO 9241-210:2010: human-centered design process for interactive systems[1]) and its goal is to design a product or service that can be intuitively and conveniently used by human users – or, in other words, to optimize a product or service around its users.[18] Despite its success in other fields, so far UCD has barely been used to design incentives in HCSs. This could be because there exist many types of HCSs with very different characteristics, and as such it is impossible to follow a set of common standards to design incentives suitable for each type. It could also be due to the fact that UCD is, in many ways, an empirical process. This means that it cannot be used to theoretically predict the outcome of different incentives – instead, one must build the system in order to evaluate the incentives' ability to deliver the desired system outcome. This can be impractical.

ICD employs game theory principles to design incentives and theoretically analyze their effect on a given HCS. ICD is generally referred to as *mechanism design* in game theory literature.[19-22] The surprising efficiency of applying ICD to solve human computation problems was demonstrated in a famous competition conducted by the Defense Advanced Research Projects Agency (DARPA) in 2009. In that competition, teams had to locate ten red balloons, placed across the continental US, as fast as possible. The winning team, from the Massachusetts Institute of Technology (MIT), located all 10 balloons in less than nine hours. They achieved this feat by using ICD to design a recursive incentive mechanism that enabled them to quickly recruit a large number of volunteers.[23-25] A number of other HCSs also relied on ICD to design appropriate incentives, such as games with a purpose.[26] Given its success and widespread adoption in the HCS field as well as its strong theoretical support, this survey reviews incentives in HCSs from the perspective of mechanism design.

The goal of this article is therefore to review how incentives are provided to human users in the state-of-the-art and influential HCSs, and how mechanism design can be used to support the design of such incentives. More specifically, this survey will answer the following research questions:

- How do the state-of-the-art HCSs provide incentives to human users?
  - What incentive mechanisms can be derived from existing HCSs?
  - What classic incentive mechanisms have been applied in HCSs?
- Where is the field of incentive design for HCSs headed in the future?

The rest of this survey is organized as follows. In Section 2, a review of eight state-of-the-art HCSs is provided, aiming at better understanding how incentives are provided in these systems. Section 3 surveys the mechanisms that have been derived from and applied to HCSs so as to provide incentives to users. Section 4 discusses research trends and challenges in the design of incentives for HCSs while Section 5 concludes this survey.

[1] http://www.iso.org/iso/catalogue_detail.htm?csnumber=52075

## 2 Typical Human Computation Systems

In this section, we provide a brief survey of state-of-the-art HCSs that have had a significant impact on society and in academia. Our goal here is not to cover all the existing HCSs, but rather to provide a thematic overview of the most interesting systems. We will present these HCSs according to their application domains. Specifically, we review four types of HCSs: metadata collection systems, question and answer websites, encyclopedia websites, and crowdsourcing e-market platforms. We focus our analysis on how these systems provide incentives for human users.

### 2.1 Metadata Collection Systems

Metadata is defined by the Oxford dictionary as "a set of data that describes and gives information about other data." [2] Examples of metadata include labels for images and audio, library catalogues, file directories, etc. In this section, we first introduce three human computation games (HCGs), also called *games with a purpose*. HCGs are interactive games whose primary goal is to collect metadata. They exploit the fun aspect of gaming as an incentive to encourage users' participation and/or efforts. The fourth metadata collection system we introduce is reCAPTCHA, a HCS built to collect metadata about scanned characters. It achieves its purpose by forcing users to recognize the presented characters before being allowed to access online contents. The common characteristic of these HCSs is that the computation (i.e. data post-processing) is a by-product of users participating in these systems.

#### 2.1.1 The ESP Game

The ESP game was developed by Luis von Ahn's research group in 2008, under the GWAP (Games with a Purpose) project at Carnegie Mellon University. The concept was initially proposed by von Ahn in 2004.[27] In this game, two players are randomly paired. They are not told who the other person is and are not allowed to communicate with each other at any point. The only common thing that the two players share is an image for which they are going to provide labels. Their goal is to enter the same label for a given image, which is referred to as "agreeing on an image". An example is shown in Fig. 2. Once both players have typed the same label, they move on to the next image. Pairs strive to agree on as many images as they can within 150 seconds. For every agreed-upon image, they get a number of points. The players can also choose to skip difficult images by both clicking the *pass* button.[27]

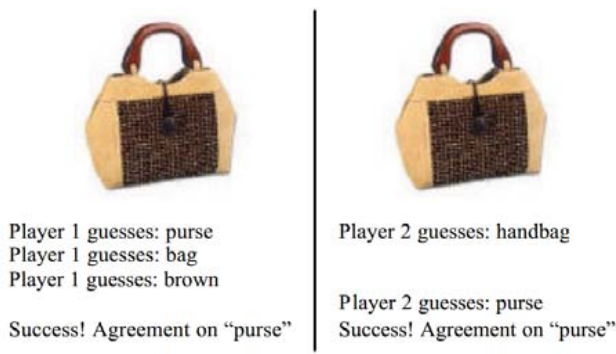

Player 1 guesses: purse
Player 1 guesses: bag
Player 1 guesses: brown

Success! Agreement on "purse"

Player 2 guesses: handbag

Player 2 guesses: purse

Success! Agreement on "purse"

**Figure 2.** Partners agree on an image in the ESP game[27]

To increase the difficulty, a list of taboo words is given – players are not allowed to enter them as guesses. These words will usually be the most frequent labels for a given image, making the game harder (and allowing the HCS to collect more interesting labels). While playing the game, the players have the incentive to describe each image using appropriate words so as to get as many points as possible.

#### 2.1.2 TagATune

TagATune was also proposed by von Ahn's research group, in 2007.[28] It is an audio-based game whose purpose is to gather perceptually meaningful metadata for audio.[28, 29] Despite ESP's success at collecting tags for images, its approach is impractical for labelling music and audio clips. The main reason is that it is very difficult for two players to agree on a description or label, due to the abstract nature of music.[29] TagAtune was thus developed to overcome this difficulty.

In a TagATune game, two randomly selected players are presented with either the same or different 30-second audio clips and are asked to type descriptions for them. They then have to guess whether they are listening to the same audio clip. After both players enter their guess, the game reveals the result of the round to the players and proceeds with the next round. In each round, the two players win points only if they both guess correctly. An example is shown on the left screenshot of Fig. 3.A game lasts three minutes in total. When the game is completed, a scoreboard displays the final score and the player's current level, as shown in the right screenshot of Fig. 3. A leaderboard is displayed on the left of the main game panel throughout the game.[29] The scoring system as well as the provided leaderboard provide players with incentives to generate accurate labels for the audio clips.

[2]http://www.oxforddictionaries.com/definition/english/metadata

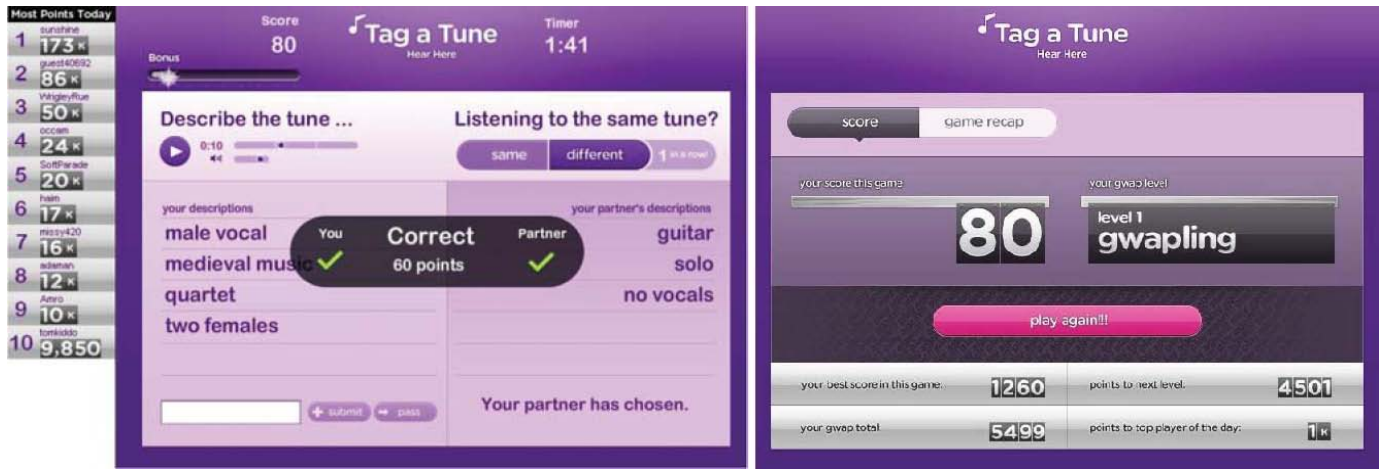

**Figure 3.** TagATune's interface[29]

*2.1.3 Verbosity*

Verbosity is another game proposed by von Ahn's research group, in 2006.[30] It is an interactive and enjoyable computer game used to collect common-sense knowledge. This knowledge consists of basic facts that a majority of humans accept as truth, such as "a basketball is round", or "fire is hot".

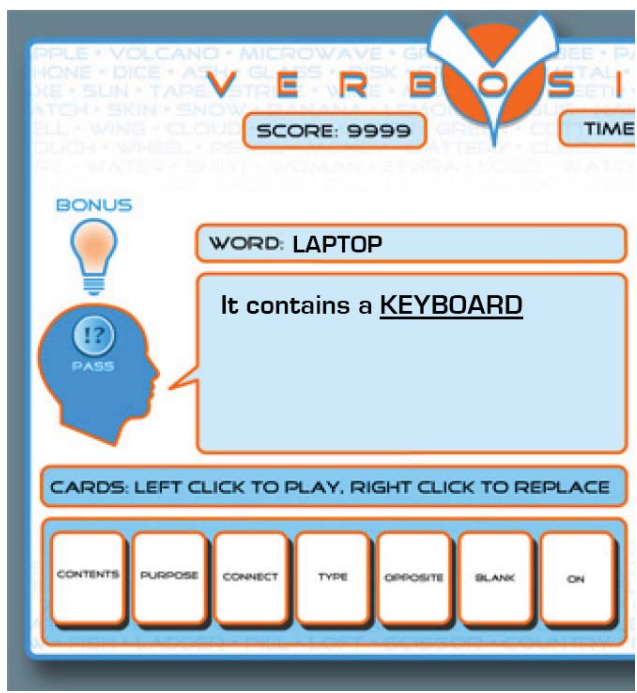

**Figure 4.** An example of narrator in Verbosity[30]

In a Verbosity game, two players are randomly selected online, and one of them is assigned as the "narrator" while the other one is the "guesser". The narrator is shown a secret word and tries to describe this word by typing related hints (not containing the secret word) using the provided sentence templates. For example, in Fig. 4, the narrator is shown with the word "LAPTOP" and types the hint "it has a KEYBOARD". The narrator has many sentence templates to choose from, which adds fun to the game. The guesser, given the hints, can type his/her guesses. The two players win points whenever the guesser correctly types out the secret word. Players take turns in narrating and guessing, and each game lasts six minutes. The narrator has the incentive to describe the words using common sense facts so that the guesser can identify the secret words as quickly as possible, thus ensuring both players get as many points as possible.

*2.1.4 reCAPTCHA*

CAPTCHAs (Completely Automated Public Turing test to tell Computers and Humans Apart) are a widespread security measure on the World Wide Web that prevents automated programs from abusing online services. They do so by asking humans to perform a task that computers cannot yet perform with acceptable accuracy, such as deciphering distorted characters. reCAPTCHA is an extension of CAPTCHA. The purpose of this HCS is to help digitize old printed materials by asking users to decipher scanned words from books that computerized optical character recognition (OCR) failed to recognize. It has been shown that this method can transcribe text with a word accuracy exceeding 99%, matching the guarantee of professional human transcribers. So far, it has been deployed in more than 40,000 websites and has transcribed over 440 million words.[31]

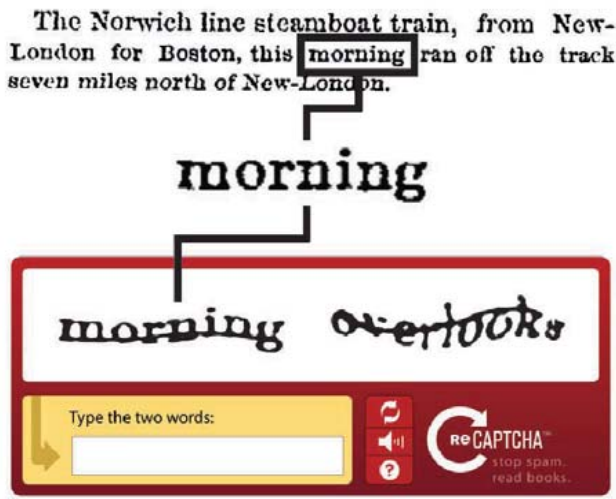

**Figure 5.** A reCAPTCHA example[31]

In reCAPTCHA, users are asked to type two words correctly before being allowed to access desired online contents, as shown in Fig. 5. One word (e.g., "morning" in Fig. 5) is the scanned word that was not recognized correctly by OCR programs – it is referred to as the *unknown word*. The other word (e.g., "overlooks" in Fig. 5) is the *control word* that was recognized correctly by OCR programs. Both words are distorted in the same way, for example by adding a line over the word. If a user wrongly types the control word, the system will ask him/her to type another pair of words until the control word is typed correctly. If users correctly type the control word, the system considers that they are human and assumes that they also typed the unknown word correctly. The main incentive of users to participe in reCAPTCHA is their need to access the online content that will come after the reCAPTCHA verification.

## 2.2 Question and Answer Websites

Question and answer (Q&A) websites are collaborative knowledge creation platforms in which reputation and social status are employed to incentivize human users to contribute valuable answers to the questions asked by other users. Popular Q&A websites include Yahoo! Answers (https://answers.yahoo.com/), Baidu Knows (http://iknow.baidu.com),[32–34] Quora (https://www.quora.com/), and Stack Overflow (http://stackoverflow.com/). In this section, we review two representative HCSs from this class: Yahoo! Answer and Stack Overflow.

### 2.2.1 Yahoo! Answers

Yahoo! Answers was launched on June 28, 2005 to replace Ask Yahoo! (Yahoo!'s former Q&A platform which was eventually discontinued in March 2006). In Yahoo! Answers, users are allowed to both submit questions they seek an answer to and answer questions asked by other users. The system uses a point and level system to provide incentives for users to participate and provide their best effort when answering questions. In this system, a newly registered user is awarded 100 points when signing up. Users are then awarded 1 point for logging in the system each day, and given 5 points for answering a question. When one of their answers is selected as the best answer to a question, they receive an additional 10 points. Asking a question costs 5 points but 3 points can be earned back by picking a best answer for that question. Based on the accumulated number of points, users are assigned one of seven "level" designations. A higher level confers greater privileges, such as the possibility to ask or answer more questions per day. In addition to these privileges, leaderboards and top contributor designations also encourage users to accumulate more points.

### 2.2.2 Stack Overflow

In 2008, Jeff Atwood and Joel Spolsky launched Stack Overflow (http://stackoverflow.com/), a free online site where both professional and amateur programmers contribute to and benefit from a library of high-quality programming-related questions and answers. Stack Overflow was further expanded to include other topics. The expanded network, called Stack Exchange (http://stackexchange.com/), comprises 130 topic-specific communities (one of them being Stack Overflow). Since then, millions of people have jumped at the opportunity to help a stranger – there are currently over 64 million monthly unique visitors in the Stack Exchange network http://stackexchange.com/about/about. Without loss of generality, we introduce the incentives provided to Stack Overflow users, which are the same as those offered to other Stack Exchange communities.

Incentives in Stack Overflow are primarily provided through a reputation system. Reputation is used to measure how much the system trusts the user. There exist a number of rules to calculate a user's reputation. For example, a user earns 5 reputation points when one of his/her questions is voted up and loses 2 reputation points when the question is voted down. Other rules can be found at http://stackoverflow.com/help/whats-reputation. The higher a user's reputation, the more privileges he/she has. Thus, users have the incentive to provide their best efforts and make valuable contributions to the system so as to boost their reputation. Besides reputation, users can also receive badges when they make especially significant contributions. Badges are classified into three levels (gold, silver and bronze), and each level contains various badges that reward specific achievements. For example, "Famous Question", one of the gold badges, is awarded to a user if one of his/her questions has 10,000+ views. Badges incentivize users to make efforts in specific aspects of the system.

## 2.3 Encyclopedia Websites

Collaboration-based free encyclopedia websites gather individuals around the world, encourage them to contribute their knowledge, and promote volunteer editorial services. Their goal is to create a high-quality, freely-accessible information resource. Popular HCSs of this type include Wikipedia (http://en.wikipedia.org/), Appropedia (http://www.appropedia.org/) and Baidu Baike (http://baike.baidu.com/). We review Wikipedia as their representative.

### 2.3.1 Wikipedia

Launched in 2001, Wikipedia is a Web platform that hosts a free online encyclopedia. Any user with an Internet connection can voluntarily edit the content at any time, without the need to register or to apply for editorial privileges.[35] Users may also choose to create a Wikipedia user account. Registered users enjoy more privileges, such as the ability to edit semi-protected content as well as the possibility to build their reputation and receive awards. It is thus recommended to create a user account, but contributions from unregistered users are also welcome.

There are three core content policies that control the type and quality of material that is acceptable in Wikipedia: "neutral point of view" (NPOV)', "verifiability", and "no original research" (NOR). Editing from a NPOV means that the content edited should be presented in a fair way, without bias[3]. Verifiability means that the information source of the content shown in Wikipedia should be checkable through the provided reliable source[4]. The third policy indicates that the content should only be based on previously published information rather than the beliefs, experiences or original research of its editors[5]. The success of Wikipedia is generally attributed to the involvement of a large number of contributors.[36] To encourage users to actively contribute to Wikipedia, incentives are provided in the form of reputation and awards. There is no formal mechanism to measure a user's reputation – it rather comes from the community's perception of a particular user's work. To take on additional administrative responsibilities (which come with privileges e.g. rollback capability, ability to delete pages, etc.), a user must ask the Wikipedia community for approval. It is therefore in a user's best interest to maintain a high reputation so that he/she may gain new responsibilities and privileges. Moreover, various awards are provided to users to recognize their contribution, such as Barnstars,[37] Project Awards, Personal User Awards, and the Editor of the Week Award.

## 2.4 Crowdsourcing E-market Platforms

Crowdsourcing e-market platforms are HCSs whose primary purpose is to recruit human users to perform different tasks. A number of such systems have been developed in the recent decades, including Amazon Mechanical Turk (http://www.mturk.com), Baidu Crowdsourcing (http://test.baidu.com/), Naver Knowledge iN (http://www.naver.com/), and Taskcn (http://www.taskcn.com/). We introduce Amazon Mechanical Turk as their representative.

---

[3]http://en.wikipedia.org/wiki/Wikipedia:Neutral_point_of_view [4]http://en.wikipedia.org/wiki/Wikipedia:Verifiability
[5]http://en.wikipedia.org/wiki/Wikipedia:No_original_research

### 2.4.1 Amazon Mechanical Turk

Launched in 2005 by Amazon.com, Inc., Amazon Mechanical Turk (MTurk) is an online marketplace for getting work done by human users (also called 'workers').[38] There were 100,000 workers from over 100 countries by March 2007, and over 500,000 workers around the world by January 2011[6]. In MTurk, individuals register as "task requesters" (shown on the right hand side of Fig. 6) or "workers" (shown on the left hand side of Fig. 6). Requesters first create and test tasks they wish workers to participate in (called human intelligent tasks – HITs) and subsequently post them. Workers can browse the available HITs and select those they wish to contribute to. They then do the required work and ultimately get paid for successfully completed tasks. The system workflow is summarized in Fig. 7. Workers are generally young, with low income and small families, and their incentive to participate is primarily linked to the monetary reward provided by the requesters.[39–4][1]

Requesters

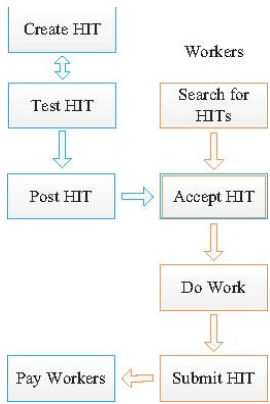

**Figure 7.** MTurk's workflow

# 3 Mechanism Design in Human Computation Systems

As mentioned previously (and as was demonstrated through the state-of-the-art HCSs described in Section 2), human users play an important role in a HCS. Providing them with appropriate incentives to encourage participation and elicit high efforts is essential for the success of any HCS. Mechanism design is a helpful tool to analyze and design such incentives. Indeed, the incentives of human users in their strategic interactions within a given system is at the core of this field.[42]

Before diving deeper into mechanism design, let us first introduce some important game theory terminology and concepts. A *mechanism* is a mathematical model that describes the interactions between human users in a game format. These users are called the *players*, and their actions are determined by the *strategy* they choose. Strategy selection is motivated by every player's *utility function*, or preference over all the possible system outcomes. A mechanism's solution (or *equilibrium*) is composed of the strategies chosen by every player in the mechanism, and thus is the outcome of the game. A mechanism may possess multiple equilibria, as players adapt their strategy depending on other players' utility functions and chosen strategies. One of the most famous solution concepts is the *Nash equilibrium*, in which no player has the intention to unilaterally change his/her strategy as he/she cannot achieve better utility by doing so. An *incentive mechanism* is a mechanism whose equilibrium helps achieve one of the system designer's goals, such as maximizing the total welfare of the system or obtaining the truthful information from the users.

Because most HCSs can be modelled as games, they can be naturally connected to mechanism design. On one hand, a mechanism can be derived by capturing a HCS's features (e.g., the users' utility functions, their actions, and their strategies). Through game theoretical analysis of the mechanism, the equilibrium strategies can be obtained. If the derived mechanism turns out to be an incentive mechanism, it will enrich the incentive mechanism literature, and it could potentially be applied in different contexts. The equilibrium strategies can also provide an explanation for the observed phenomena in existing HCSs, i.e. why users behave in a particular manner. On the other hand, classic mechanisms in the literature of mechanism design can also be applied in the design of HCSs so as to achieve the objective of HCS designers. As will be discussed later, this has been done successfully in a number of existing systems.

The dual relationship between HCSs and mechanism design is shown in Fig. 8.

[6]http://en.wikipedia.org/wiki/Amazon_Mechanical_Turk

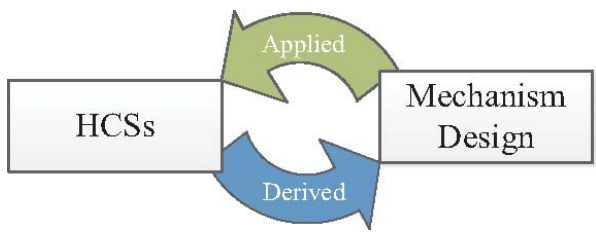

**Figure 8.** The relationship between HCSs and mechanism design

In subsection 3.1, we review the mechanisms derived from HCSs and the corresponding game theoretical analysis, and in subsection 3.2, we review the classic mechanisms that have been successfully applied in HCSs.

## 3.1 Mechanisms Derived from HCSs

The state-of-the-art HCSs introduced in Section 2 all provide incentives for users to participate in the system or to provide their best efforts, and a mechanism may be derived from each of them. However, the literature only covers five derived mechanisms. We review them below.

### 3.1.1 Output-agreement mechanism

The output-agreement mechanism[26, 43] is derived from the ESP game, and it is presented in Fig. 9.

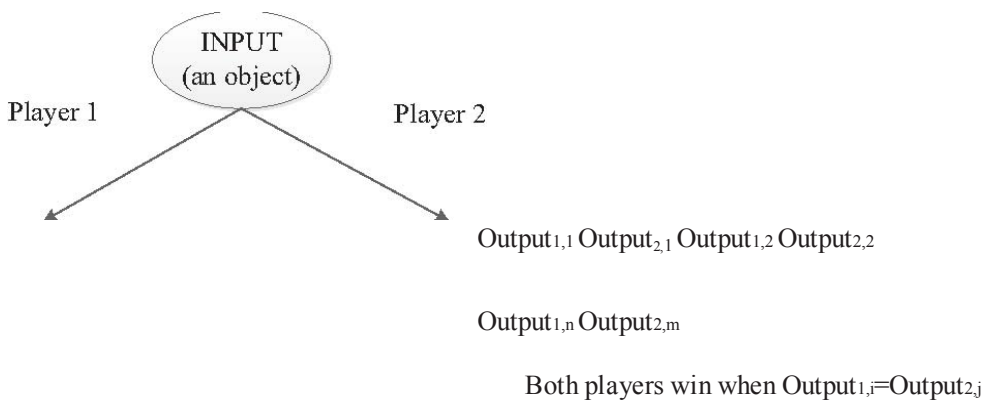

**Figure 9.** The output-agreement mechanism[26]

The output-agreement mechanism can be described as a two-stage game model with imperfect information.[44] In the first stage, a player $i$ is presented with an object and creates a dictionary of outputs to represent the object. In this stage, the strategy of the player is binary: $e_i \in \{L,H\}$ where $L$ represents *low effort* and $H$ represents *high effort*. The size of the output dictionary under low effort is smaller than the size of the dictionary under high effort (i.e. high effort leads to a richer dictionary). In the second stage, the player makes a decision $s_i$ about the order in which to sort the chosen output dictionary – this sorting will determine the order in which the words will be entered/revealed in the game. The strategy of the player in the second stage is also binary: $s_{ii}^{\uparrow}$ is the strategy of sorting the words in the order of decreasing frequency (popularity) and $s_{ii}$ is the strategy of sorting them by increasing frequency. In other words, adopting the $s_i$ strategy implies that a player will start by revealing the most common words in his/her dictionary and will leave the rarer words to the end. A complete strategy for the player in the mechanism is thus defined by $(e_i, s_i)$. The outcome of the game is modelled as $o = (w, l)$, where $w$ is the first agreed-upon word and $l$ is the location (rank) of that word in the two players' ordered dictionaries. If this rank differs between both players, $l$ is taken as the largest value of the two. The utility function corresponds to the players' preference over the possible outcomes. In this mechanism, both players have the same preference: *match-early preference*. Both players prefer to match their outputs as fast as possible i.e. they want the smallest $l$ value possible.

It has been shown in[44] that the symmetric equilibrium $(L, s_i^{\downarrow})$ is a Nash equilibrium in which the expected utility of a player is maximized given that the other player also adopts the same strategy.

---

[7]In a symmetric equilibrium, every player has the same equilibrium strategy.

In other words, in the output-agreement mechanism with match-early preference, players focus their attention on creating a small dictionary and reveal high-frequency (common) outputs first. Thus, it is not surprising to observe that in practice, ESP players provide the most common tags, even if these tags do not describe the shown image well.[27, 44] This is a problem since the gathered metadata is not as useful and descriptive as it could be.

In order to solve this problem, a new preference model was proposed in.[44] This new model is called *rare-output preference*, i.e. players achieve better utility by matching on infrequent outputs. In the output-agreement mechanism with rare-output preference, the Nash equilibrium strategy of a player $i$ is $(H, \overset{\uparrow}{s_i})$. Thus, the output-agreement mechanism with rare-output preference is an incentive mechanism for promoting high effort and low frequency outputs.

### 3.1.2 Input-agreement Mechanism

The input-agreement mechanism[26] is derived from the TagATune game, and it is shown in Fig. 10.

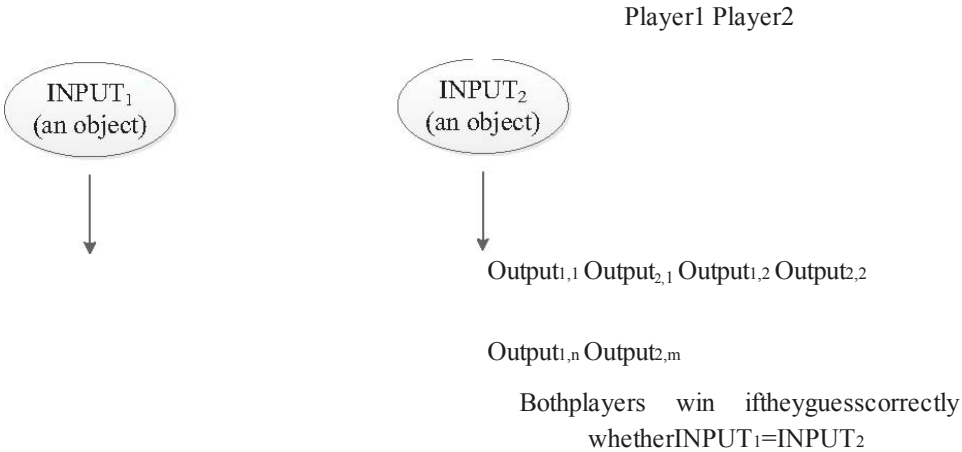

Player1 Player2

$INPUT_1$ (an object)   $INPUT_2$ (an object)

$Output_{1,1}$ $Output_{2,1}$ $Output_{1,2}$ $Output_{2,2}$

$Output_{1,n}$ $Output_{2,m}$

Both players win if they guess correctly whether $INPUT_1 = INPUT_2$

**Figure 10.** The input-agreement mechanism[26]

The outcome of the mechanism is not directly determined by what outputs the players generate for the object, but by both players' guess whether they are shown the same input object. Before a game starts, neither player holds the correct answer (i.e. whether they are shown the same input object), and each player has to derive his/her guess from the other player's outputs during the game. The players only get rewards if they both guess correctly. As a result, in this mechanism, the strategy that maximizes players' utility is to describe the given input objects using informative outputs, which requires high effort. Since the system objective is to collect informative/descriptive metadata for the input objects, the input-agreement mechanism is an incentive mechanism.

### 3.1.3 Inversion-problem Mechanism

The inversion-problem mechanism[26] is derived from the Verbosity game, and it is shown in Fig. 11.

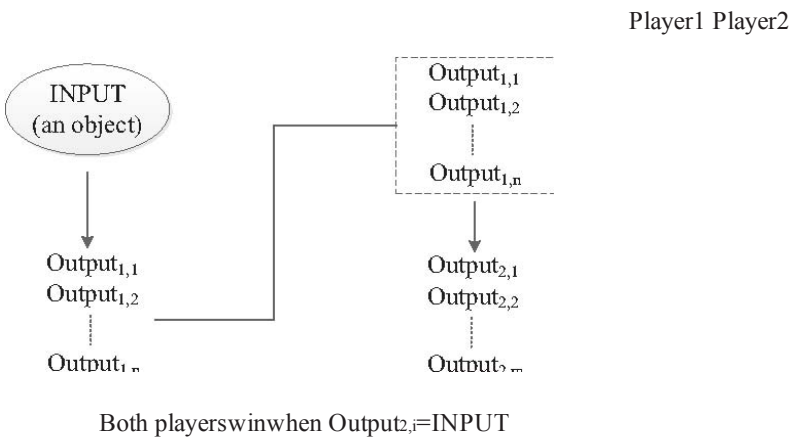

Player1 Player2

INPUT (an object)

$Output_{1,1}$ $Output_{1,2}$

$Output_{1,n}$

$Output_{1,1}$ $Output_{1,2}$

$Output_{1,n}$

$Output_{2,1}$ $Output_{2,2}$

$Output_{2,m}$

Both players win when $Output_{2,i} = INPUT$

**Figure 11.** The inversion-problem mechanism[26]

This mechanism is different from the output-agreement and input-agreement mechanisms, as it uses an asymmetric two-player game model. One player is given an input object and provides outputs to describe it. The other player does not have access to the input object and instead has to guess what it is by generating outputs based on the outputs given by the first player.

The first player's outputs must not contain the input object. The two players get rewards only when the second player correctly guesses what the input object is. The unique Nash equilibrium is that the first player describes the input using common knowledge (so that it is easier for the second player to guess what it is) while the second player makes his/her best effort to guess the input as early as possible. Since the system objective is to collect common knowledge about the input object, the inversion-problem mechanism is an incentive mechanism.

### 3.1.4 Rewarding Best-answer Mechanism

The rewarding best-answer mechanism[45] was derived from Yahoo! Answers. In this mechanism, an *asker* asks a question, and users provide answers. The answers to a question are considered to be provided sequentially. The mechanism assumes that it is costless for users to answer a question and that every user has the ability to combine his/her own answer with those previously provided by other users. An asker closes his/her question (that is, prevents any other users from answering the question) when the collected information reaches a private threshold that follows a uniform distribution. Upon closing a question, the asker selects one answer as the best answer, hence the name of this mechanism. The user that provided the best answer is awarded some points. Depending on the nature of the information contained in the answers to a question, two cases must be separately analyzed. In the *complement information* case, the value of two pieces of information that are combined is higher than the sum of the values of each piece. In the *substitute information* case, the value of the second piece of information is smaller than the value of the first piece of information.[46]

Through theoretical analysis, it can been shown that in the complement information case, the unique pure-strategy Nash equilibrium of the mechanism is that all users delay their participation as much as possible, which is the least efficient solution. This result helps explain the commonly observed phenomenon that many users delay their answers until the last second in a number of online question and answer forums. In the substitute information case, the unique pure-strategy Nash equilibrium is that all users participate as quickly as possible, which is the most efficient solution.[45] In this case, the rewarding best-answer mechanism is an incentive mechanism that promotes early participation.

### 3.1.5 Virtual Badge Design Mechanisms

In many HCSs, a system of *virtual badges* is used to recognize a user's contributions to the site. For example, the "Popular Question" badge is awarded to a user when one of his/her questions is viewed more than 1000 times in Stack Overflow. Badges have been shown to be highly valued by users, who actively pursue and compete for them.[47] Furthermore, users generally increase their efforts when they are close to receiving a badge.[48]

Through an analysis of the existing HCSs, two types of virtual badge design mechanisms have been derived: the absolute virtual badge design mechanism and the relative virtual badge design mechanism. In the case of the absolute virtual badge design mechanism $M_\alpha$, badges are awarded to users whose overall contribution meets a predetermined standard $\alpha$. This mechanism is used in Stack Overflow and in many more HCSs. In the relative virtual badge design mechanism $M_\rho$, badges are awarded to a predetermined percentage $\rho$ of top contributors. This mechanism is deployed in MTurk for instance. Both mechanisms possess equilibria.[49] Specifically, $M_\alpha$ has zero-participation equilibria for standards that are too high, whereas all equilibria of $M_\rho$ have non-zero participation for all possible $\rho$. In other words, the absolute virtual badge design mechanism with low standard and the relative virtual badge design mechanism are incentive mechanisms to promote non-zero participation.

The analysis in[49] has shown that the curvature of the value function of winning a badge determines whether the number of winners of a particular badge should be revealed to the users. More specifically, if the value function is strictly convex, making the number of winners publicly known can increase equilibrium effort; if the value function is strictly concave, publicly revealing the number of winners reduces equilibrium effort.

### 3.1.6 Summary

In the previous subsections, we introduced eight mechanisms derived from six HCSs. These mechanisms are summarized in Table 1. Analysing the derived mechanisms through the mathematical lens provided by mechanism design enables us to explain some of the observed phenomena in existing HCSs (i.e. why users behave in certain ways). Each discussed HCS possesses an incentive mechanism that achieves one of two system objectives: promoting high effort from human users or attracting user participation. However, no incentive mechanism covers both objectives simultaneously, which is a limitation.

## 3.2 Classic Mechanisms That Have Been Applied in Human Computation Systems

In the literature of mechanism design, many classic mechanisms were proposed to achieve the objectives of system designers, including promoting efforts from users and attracting user participation. In this section, the classic mechanisms that have been applied to HCSs are reviewed. We will point out the incentives they provide and the HCSs in which they have been applied.

**Table 1.** A summary of the mechanisms derived from the HCSs

| HCS | Mechanism | Equilibrium Strategy | | Incentive Mechanism? |
|---|---|---|---|---|
| | | Effort | Participation | |
| ESP | Output-agreement with match-early preference Output-agreement with rare-output preference | Low High | — — | √ |
| TagATune | Input-agreement | High | — | √ |
| Verbosity | Inversion-problem | High | — | √ |
| Yahoo! Answers | Rewarding Best-answer in complement case Rewarding Best-answer in substitute case | — — | Early Late | √ |
| Stack Overflow | Absolute Virtual Badge Design with high standard Absolute Virtual Badge Design with low standard | — — | Zero Non-zero | √ |
| MTurk | Relative Virtual Badge Design | — | Non-zero | √ |

### 3.2.1 Direct Revelation Mechanisms

The revelation principle is a technical insight that allows mechanism designers to make a general statement about what mechanisms are feasible subject to incentive constraints.[50] The revelation principle tells us that for any general indirect mechanism (in which the private information held by players is not directly revealed) with a Nash equilibrium, there exists an equivalent incentive-compatible direct revelation mechanism in which truthful revelation is an equilibrium.[51] With the support of the revelation principle, without loss of generality, it is possible to restrict the message space to the observed information space and focus our attention on direct revelation mechanisms.

In this mechanism, there are $n$ players. Each player $i$ possesses private information $t_i$ which is unknown to the mechanism or to other players. He/she is required to report his/her information. The reported information $\theta_i$ may or may not be different from the private information, i.e. $t_i = \theta_i$ or $t_i = \theta_i$. By aggregating the reported information from every player, an outcome is determined by the mechanism and is denoted by $o(\theta_1,...,\theta_n)$. The utility function of player $i$ is given by $u(t_i,o)$. The mechanism is a direct revelation mechanism if $u(t_i,o) \geq u(\theta_i,o)$ for all $\theta_i$. Thus, the strategy of making high effort through providing truthful information is an equilibrium strategy.

Direct revelation mechanisms have been applied in HCSs to incentivize users to make high effort in truthfully revealing information they possess. In the TagATune and Verbosity games, the equilibrium strategy is for users to make high efforts when describing the provided input object (audio and text, respectively). In reCAPTCHA, users have to directly and correctly type the scanned characters before they can access desired online resources.[31] In MTurk, the payment scheme of some tasks is maximized for users that directly reveal their truthful answer.[52]

### 3.2.2 Majority Voting Mechanism

A voting system counts the votes from voters and aggregates them to yield a final result (including a winner). The study of formally defined voting systems is called social choice theory or voting theory, and it is a subfield of mechanism design.[17,53] The winner can be a single user or multiple users. Without loss of generality, we introduce the voting mechanism for the single winner case.

In the majority voting mechanism, one of $n$ candidates is to be selected as the winner. Each voter (or player) has a (possibly incomplete) preference ranking over the set of candidates according to his/her utility function and votes accordingly. The candidate that receives the largest number of votes becomes the winner.[54] The incentive of players participating in the mechanism comes from the link between their utility and the voting outcome. For example, people in a community would vote for a highly qualified senator because he can represent them well and will defend their interests. If the set of candidates is also the set of voters, each voter then has the incentive to become the winner and will make high effort.

The majority voting mechanism has been naturally applied in many HCSs in which the ratings of users are used to compute the quality of online content. For example, some virtual badge winners in Stack Overflow and Yahoo! Answer are selected through this mechanism.

### 3.2.3 Recursive or Referral Incentive Mechanism

The recursive or referral incentive mechanism has been commonly used in a diffusion-based task environment, in which players become aware of tasks either by being directly informed by the mechanism or by being indirectly informed by other players connected with them.[25] When a task is completed, the mechanism is able to identify both the player who executed it and the connection pathway that led to that player. The mechanism then provides rewards to all players in the connection pathway in a recursive or referral way. The objective is to attract users and encourage them to participate.

This type of mechanism has been used by many online websites, for example Groupon and Dropbox. Its effectiveness in recruiting users has been empirically demonstrated in.[55] Even though it has not been used in the reviewed HCSs, it was effectively applied in addressing a human computation problem that we briefly discussed in the introduction. In 2009, the Defense Advanced Research Projects Agency (DARPA) challenged teams to locate ten red balloons located across the continental USA as quickly as possible and offered $40,000 in prize to the winning team. An MIT team designed a recursive incentive mechanism to tackle the challenge. If they won the competition, they promised to use part of the prize money as a financial incentive to reward the people who correctly located balloons as well as those that connected the finder with the MIT team in the first place. For each balloon found, the team would provide up to $4,000: $2,000 were promised to the person locating the balloon, $1,000 to the person who recruited that balloon finder onto the team, $500 to whoever recruited the recruiter, $250 to whoever recruited that person, and so on. Using this mechanism, the MIT team successfully located all ten balloon in less than nine hours and won the competition.

### 3.2.4 All-Pay Auction Mechanism

In an all-pay auction mechanism involving a single indivisible item, a set of bidders share a common objective – they each want to win the item. Each bidder $i$ has a private value $v_i$ towards the item and bids for it by submitting an envelope containing an amount of money $x_i$ they are willing to pay for the item. The item is won by the bidder with the highest $x_i$. All of the bids submitted by bidders are forfeited, i.e. nobody gets their money back no matter the outcome of the auction. The all-pay auction mechanism has been widely employed in the contest literature because it captures the essential features of a competition environment.[56,57] Crowdsourcing contests (e.g. Wikipedia, Yahoo! Answer, Baidu Baike and Stack Overflow) can also be treated as an implementation of the all-pay auction mechanism – they achieve their objective of eliciting high quality contributions through rewarding the top contributors.[58–60] In order to increase their chances of being rewarded, users have the incentive to provide their maximum "bid" in the form of making their best contribution. The relationship between the rewards and the participation level was studied in.[57] They showed that the participation level linearly increases with the logarithm of the reward amount.

### 3.2.5 Summary

The classic mechanisms that have been used in HCSs are summarized in Table 2. We include the incentives they provide as well as the HCSs to which they are applied. Among the four mechanisms reviewed, the direct revelation mechanism provides incentives for users to make their best effort in revealing the observed information, and the other three mechanisms incentivize both user participation and effort.

**Table 2.** A summary of the classic mechanisms applied in HCSs

erbosity $_{TagATune/VeCAPTCHA}$

Yahoo!

Answer $_{StackOverflow}$
MTurkikipedia

| Providing Incentives For | Mechanisms | r | | MTurk | Wikipedia | W | |
|---|---|---|---|---|---|---|---|
| Efforts | Direct Revelation | √ | √ | | | | √ |
| Participation | Majority Voting | | | √ | √ | | |
| + | Recursive or Referral | | | | | | |
| Efforts | All-Pay Auction | | | √ | √ | √ | |

## 4 Future Research Directions

In this section, future research directions are discussed. We separate them into two broad categories: 1) further developing our understanding of human users, and 2) further developing mechanism design. Both of these aim to facilitate the design of better incentives for HCSs.

## 4.1 Further Developing our Understanding of Human Users

In order to promote user participation and elicit high efforts in a HCS, it is essential to have a deep understanding of the target users. For this purpose, four specific directions appear promising: 1) using theories of needs and motivation to better understand how users behave, studying how 2) demographics and 3) context affect user behaviour, and 4) analyzing the feasibility of incentives from a user's perspective.

### 4.1.1 Incorporating Theories of Human Needs and Motivation

The concept of incentive that we discussed in this survey is closely related with the notion of motivation found in psychology. Motivation theories can thus be helpful to better understand why humans act in certain ways, and how incentives may encourage different behaviours. Human needs are generally recognized by psychologists and behavioural scientists as the source of the motivations that drive behaviours.[61-65] In,[62] Maslow defined five essential needs: physiological, safety, love/belonging, esteem, and self-actualization. McClelland identified three motivators that he believed all humans have: a need for achievement, a need for affiliation, and a need for power.[61] Moreover, the self-determination theory, an empirically-based theory of human motivation, focuses on autonomous motivation and controlled motivation as predictors of human behaviours.[66-68]

There would be value in using such rich research literature to study the intrinsic needs and motivations of HCS users so that incentives that cater to specific needs or motivations can be designed.

### 4.1.2 Studying the Effect of Demographics

It has been widely recognized that not all incentives have the same effect on all human users.[69-71] For instance, in MTurk, it has been found that young men are more likely to fail the qualification task. Professionals, students, and non-workers are more likely to take the requested task more seriously than financial, hourly, and other workers. Men over 30 and women are more likely to answer seriously.[72] Moreover, comparing Wikipedia (which currently has 4.6 million articles[8]) and Baidu Baike (6.2 million articles[9]), we find that Baidu Baike attracts more contributors even though both HCSs utilize similar incentive mechanisms, and despite Baidu Baike having a smaller potential user base than Wikipedia.

From these two examples, we notice that demographics appear to play an important role in how users behave in HCSs. However, there currently exist few user studies that look into the effect of demographics on user behaviours. It would thus be worthwhile to conduct more demographics studies so that appropriate incentives can be designed based on the target user population. This could be done by taking advantage of human computation based e-markets, such as MTurk, as such systems allow us to cheaply and quickly collect user input.[73]

### 4.1.3 Studying the Effect of Context

Similar to demographics, context can play a role in what incentives humans respond to. Environmental, physical, social, and cognitive factors can all affect user behaviour. For example, a person is more likely to be motivated by the promise of a food reward when he/she is hungry, as opposed to when he/she is full. One teenage boy might be motivated to clean his room by the promise of a coveted video game, while another one would find such an incentive completely unappealing. There is value in studying how context factors are related with the incentives of human users so that these factors can be considered when designing incentives for HCSs. This way, system designers can ensure that users adopt the desired behaviours in specific contexts.

### 4.1.4 Analyzing the Feasibility of Incentives

When we design incentives in HCSs, it is important to know whether the designed incentives are feasible from the users' perspective. An incentive that is either infeasible or that is perceived as such by users will lead to users ignoring it, which means that the system objective will not be met. Feasibility can be linked to the malleability property of intelligence, in particular in the area of education.[74-76] According to the theory of intelligence, there exist two types of mindset when it comes to intelligence: the fixed mindset and the growth mindset. Individuals who hold a fixed mindset believe that they have an unchangeable amount of intelligence. Studies have shown that people with such a mindset view challenging situations as "tests" of how much intelligence they have, and view effort and mistakes as indications of low ability.[74, 76] On the other hand, individuals who hold a growth mindset believe that intelligence is malleable, and that people can increase it through hard work. They have been shown to value learning over performance, and view effort as a necessary part of the learning process.[74-76]

Both mindsets have different implications in terms of the feasibility of incentives. A fixed mindset will favour performance over effort, so low-standard incentives (or easy tasks) are preferable. On the other hand, a growth mindset will favour high effort, which is generally desirable since most HCSs' objective is to elicit high effort from users. Some incentive structures (e.g. the "brain points" structure[77]) have been shown to promote the growth mindset and can encourage effort.

Moreover, a feasible incentive can only be achieved with a feasible task. The goal-setting theory focuses on studying the relationship between the difficulty and specificity of a goal and the humans' performance. Locke's research showed that specific and difficult goals led to better task performance than vague or easy goals.[78]

Therefore, applying the malleability of intelligence and goal-setting theories offers the potential to improve the feasibility of incentives in HCSs.

### 4.2 Further Developing Mechanism Design for Human Computation Systems

As we discussed in Section 3, mechanism design has proven to be a useful tool for HCSs. However, there are still a number of open opportunities to further develop it to improve HCSs. We will discuss four ways in which mechanism design can be expanded that HCSs would benefit from: 1) modelling human preference, 2) analyzing potential risks and designing prevention schemes, 3) evaluating the performance of mechanisms, and 4) selecting and combining existing mechanisms.

#### 4.2.1 Modelling Human Preference

Modelling the preference or utility of humans is a challenging issue in HCSs. This is because the preference of human users is personalized and dynamic. As such, traditional preference models such as decision trees[79] may fail. Thus, new formalization methodologies should be explored and more research should be carried to incorporate personality and variability when modelling human users' preference in HCSs.

#### 4.2.2 Analyzing Potential Risks and Designing Prevention Schemes

Through a rigorous analysis of incentives using mechanism design, it is possible for HCS designers to obtain all the possible equilibria of the system. Some of these equilibria may be undesirable or harmful for the achievement of the objective of the designed system. However, there is always the possibility that a specific system will realize one of these bad equilibria. As such, one future research direction is to develop an HCS 'risk assessment' methodology through equilibria analysis. Furthermore, when potential risks are identified, there would be value in designing schemes that protect the systems from them.

#### 4.2.3 Evaluating the Performance of Mechanisms

It is important to objectively evaluate the performance (or effectiveness) of a mechanism in providing incentives for users in a HCS. For this purpose, a set of guidelines or metrics should be developed. Such a set does not exist currently, and few metrics have been designed. One of the few examples is the mechanism robustness against user misbehaviour metric proposed in.[80] More research should be carried out to develop new guidelines and metrics.

Also, due to the fact that HCS users are human beings, mistakes will undoubtedly occur. In mechanism design, human users are implicitly assumed to be rational, meaning that they are able to make correct decisions all the time. This assumption is not always realistic but is necessary so that theoretical analysis can be carried out. An important research direction would thus be to model and categorize different types of user misbehaviour and then reevaluate the performance of mechanisms using a robustness metric as well as other indicators.

#### 4.2.4 Selecting and Combining Mechanisms

Selecting the proper mechanism(s) for a specific HCS is an essential task. More research should be conducted to analyze the main factors determining appropriate mechanism selection. Furthermore, as we have observed in Table 2, different mechanisms can be applied in a HCS simultaneously. An interesting research problem would be to investigate the overall effect on incentives when multiple mechanisms work together. Is it an "addition", "multiplication", or "subtraction" effect?

## 5 Conclusion

In this survey, we reviewed the fast-developing field of human computation systems from the perspective of how incentives are provided to attract human participation and elicit high efforts. We then approached the incentive issue through a mechanism design lens, as it provides a strong theoretical framework to design and analyze incentives. We introduced mechanisms derived from HCSs as well as classic mechanisms applied to these systems. We finally provided eight distinct research directions for incentive design in HCSs, organised in two broad categories: better understanding human users and further developing mechanism design.

Given the current trends as well as the widespread availability and popularity of online communication technologies, it can be expected that the number of human computation systems will keep growing. System designers will keep looking for smart, creative and efficient incentive schemes to attract as many users as possible and to elicit quality contributions from them, so this field will keep attracting significant attention in the future. Only when computers are able to reason and solve problems like humans will HCSs become obsolete. HCS researchers and designers can be confident that this scenario will not materialize anytime soon and that their work will be relevant for the decades to come!